\documentclass[twocolumn,pre]{revtex4-1}
\usepackage{color}
\usepackage{graphics,graphicx,epsfig,wrapfig}
\graphicspath{.}

\usepackage{epsf,epstopdf}
\usepackage{amssymb,amsfonts,amsmath}
\usepackage{enumitem}
\usepackage{setspace}
\usepackage{longtable}

\usepackage{ifthen}
\usepackage{placeins} 

\newcommand\mydots{\hbox to 0.8em{.\hss.\hss.}}
\newcommand{\beqn}{\begin{eqnarray}}
\newcommand{\eeqn}{\end{eqnarray}}
\newcommand{\beq}{\begin{equation}}
\newcommand{\eeq}{\end{equation}}

\newcommand{\ch}{\color{black}}

\newcommand{\ff}{1}

\newcommand{\mytitle}{Longitudinal high-throughput TCR repertoire profiling reveals the dynamics of T cell memory formation after mild COVID-19 infection}

\newcommand{\myauthors}{Anastasia A. Minervina$^{1}$, Ekaterina A. Komech$^{1,2}$,
Aleksei Titov$^{3}$, Meriem Bensouda Koraichi$^{4}$, Elisa Rosati$^{5}$, Ilgar Z. Mamedov$^{1,2,6,7}$, Andre Franke$^{5}$, Grigory A. Efimov$^{3}$, Dmitriy M. Chudakov$^{1,2,6,8}$, \\  Thierry Mora$^{4}$, Aleksandra M. Walczak$^{4}$, Yuri B. Lebedev$^{1,9}$, Mikhail V. Pogorelyy$^{1,2}$}

\makeatletter
\def\@seccntformat#1{%
  \expandafter\ifx\csname c@#1\endcsname\c@section\else
  \csname the#1\endcsname\quad
  \fi}
\makeatother

\begin{document}
\title{\mytitle}
\author{\myauthors}
\affiliation{~\\
\normalsize{$^{1}$ Shemyakin-Ovchinnikov Institute of Bioorganic
  Chemistry,}
\normalsize{Moscow, Russia}\\
\normalsize{$^{2}$Pirogov Russian National Research Medical University, Moscow, Russia}\\
\normalsize{$^{3}$National Research Center for Hematology, Moscow, Russia}\\
\normalsize{$^{4}$Laboratoire de physique de l'\'Ecole normale sup\'erieure,}
\normalsize{ENS, PSL, Sorbonne Universit\'e, Universit\'e de Paris, and CNRS, Paris,
  France}\\
\normalsize{$^{5}$Institute of Clinical Molecular Biology, Kiel University, Kiel, Germany}\\
\normalsize{$^{6}$Masaryk University, Central European Institute of Technology, Brno, Czech Republic}\\
 \normalsize{$^{7}$V.I. Kulakov National Medical Research Center for Obstetrics, Gynecology and Perinatology, Moscow, Russia}\\
\normalsize{$^{8}$Center of Life Sciences, Skoltech, Moscow, Russia}\\
\normalsize{$^{9}$Moscow State University, Moscow, Russia}\\
}

\begin{abstract} 
{
COVID-19 is a global pandemic caused by the SARS-CoV-2 coronavirus. {\ch T cells play a key role in the adaptive antiviral immune response by killing infected cells and facilitating the selection of virus-specific antibodies. }
However neither the dynamics and cross-reactivity of the SARS-CoV-2-specific T cell response nor the diversity of resulting immune memory are well understood. In this study we use longitudinal high-throughput T cell receptor (TCR) sequencing to track changes in the T cell repertoire following two mild cases of COVID-19. {\ch In both donors we identified CD4+ and CD8+ T cell clones with transient clonal expansion after infection.} {\ch The antigen specificity of CD8+ TCR sequences to SARS-CoV-2 epitopes was confirmed by both MHC tetramer binding and presence in large database of SARS-CoV-2 epitope-specific TCRs.} We describe characteristic motifs in TCR sequences of COVID-19-reactive clones {\ch and show preferential occurence of these motifs in publicly available large dataset of repertoires from COVID-19 patients. }
We show that in both donors the majority of infection-reactive clonotypes acquire memory phenotypes. Certain {\ch T cell} clones were detected in the memory fraction at the pre-infection timepoint, suggesting participation of pre-existing cross-reactive memory T cells in the immune response to SARS-CoV-2.
}
\end{abstract}

\maketitle

\section*{Introduction}
COVID-19 is a global pandemic caused by the novel SARS-CoV-2 betacoronavirus \cite{vabret_immunology_2020}. T cells are crucial for clearing respiratory viral infections and providing long-term immune memory \cite{schmidt_cd8_2018,swain_expanding_2012}. Two major subsets of T cells participate in the immune response to viral infection in different ways: activated CD8+ T cells directly kill infected cells, while subpopulations of CD4+ T cells produce signaling molecules that regulate myeloid cell behaviour, drive and support CD8 response and the formation of long-term CD8 memory, and participate in the selection and affinity maturation of  antigen specific B-cells, ultimately leading to the generation of neutralizing antibodies. In SARS-1 survivors, antigen-specific memory T cells were detected up to  11 years after the initial infection, when viral-specific antibodies were undetectable \cite{ng_memory_2016,oh_engineering_2011}. The T cell response was shown to be critical for protection  in SARS-1-infected mice \cite{zhao_t_2010}. Patients with X-linked agammaglobulinemia, a genetic disorder associated with lack of B cells, have been reported to recover from symptomatic COVID-19 \cite{quinti_possible_2020,soresina_two_2020}, suggesting that in some cases T cells are sufficient for viral clearance. Theravajan et al. showed that activated CD8+HLA-DR+CD38+ T cells in a mild case of COVID-19 significantly expand following symptom onset, reaching their peak frequency of 12\% of CD8+ T cells on day 9 after symptom onset, and contract thereafter \cite{thevarajan_breadth_2020}. Given the average time of 5 days from infection to the onset of symptoms \cite{bi_epidemiology_2020}, the dynamics and magnitude of T cell response to SARS-CoV-2 is similar to that observed after immunization with live vaccines \cite{miller_human_2008}. 
SARS-CoV-2-specific T cells were detected in COVID-19 survivors by activation following stimulation with SARS-CoV-2 proteins \cite{ni_detection_2020}, or by viral protein-derived peptide pools \cite{weiskopf_phenotype_2020,braun_presence_2020,snyder_magnitude_2020,le_bert_sars-cov-2-specific_2020,meckiff_single-cell_2020,bacher_pre-existing_2020,peng_broad_2020}. Some of the T cells activated by peptide stimulation were shown to have a memory phenotype \cite{weiskopf_phenotype_2020,le_bert_sars-cov-2-specific_2020,mateus_selective_2020}, and some potentially cross-reactive CD4+ T cells were found in healthy donors \cite{braun_presence_2020,grifoni_targets_nodate,sekine_robust_2020,bacher_pre-existing_2020}. 

T cells recognise short pathogen-derived peptides presented on the cell surface of the Major Histocompatibility Complex (MHC) using hypervariable T cell receptors (TCR). TCR repertoire sequencing allows for the quantitative tracking of T cell clones in time, as they go through the expansion and contraction phases of the response. It was previously shown that quantitative longitudinal TCR sequencing is able to identify antigen-specific expanding and contracting T cells in response to yellow fever vaccination with high sensitivity and specificity  \cite{minervina_primary_2020,pogorelyy_precise_2018,dewitt_dynamics_2015}. Not only clonal expansion but also significant contraction from the peak of the response are distinctive traits of T cell clones specific to the virus \cite{pogorelyy_precise_2018}. 

In this study we use longitudinal TCRalpha and TCRbeta repertoire sequencing to quantitatively track T cell clones that significantly expand and contract {\ch after recovery from a mild COVID-19 infection}, and determine their phenotype. We reveal the dynamics and the phenotype of the memory cells formed after infection, identify pre-existing T cell memory clones participating in the response, and describe public TCR sequence motifs of SARS-CoV-2-reactive clones, suggesting a response to immunodominant epitopes.  

\section*{Results}

\subsection*{Longitudinal tracking of TCR repertoires of COVID-19 patients}

\begin{figure*}
\noindent\includegraphics[width=\ff\linewidth]{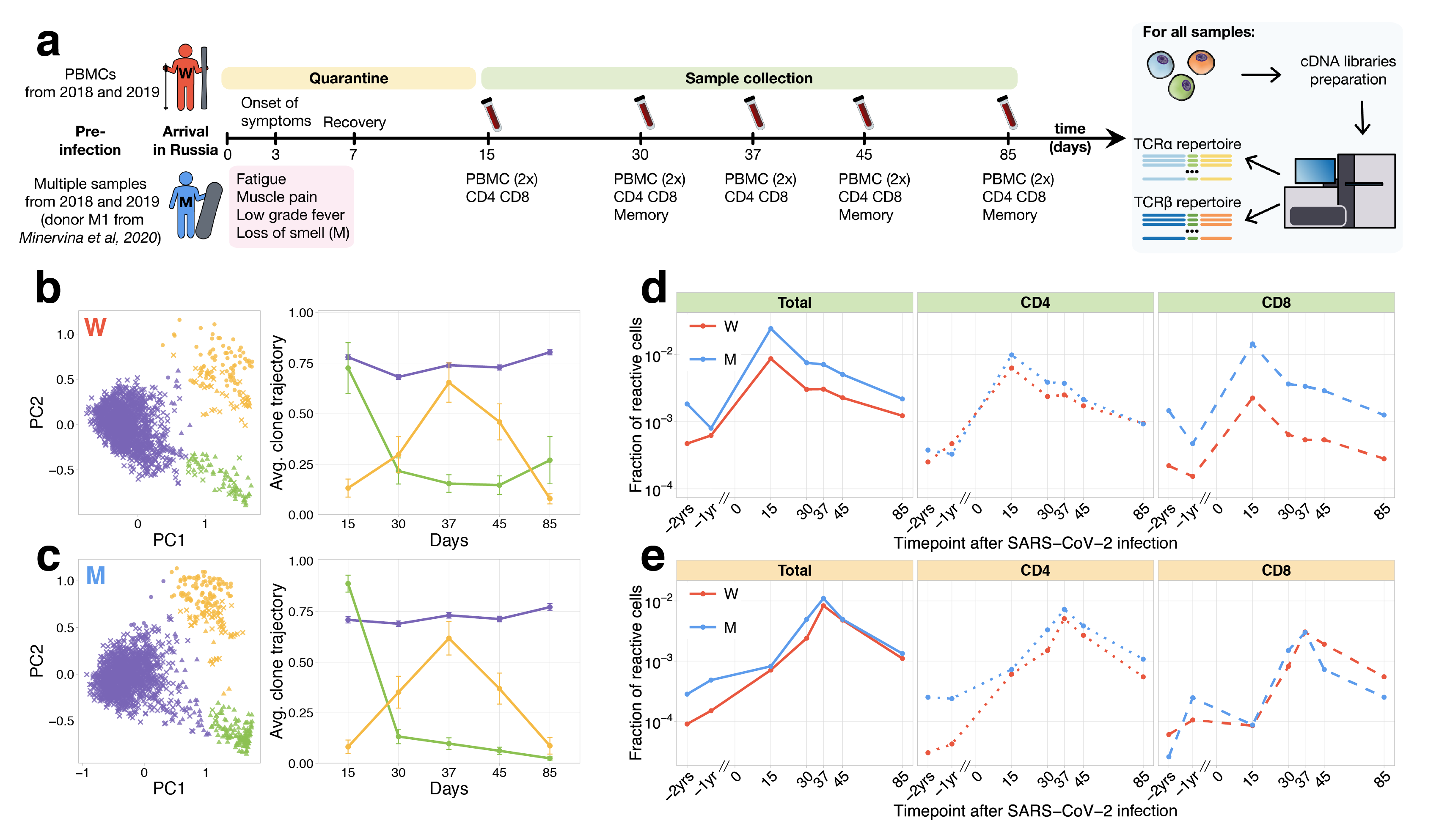}
\caption{{\bf Longitudinal tracking of T cell clones after mild COVID-19.  a, Study design.} Peripheral blood of two donors was sampled longitudinally on days 15, 30, 37, 45, {\ch 85} after arrival in Russia. At  each timepoint, we evaluated SARS-CoV-2-specific antibodies using ELISA (Fig. S1) and isolated PBMCs in two biological replicates. Additionally, CD4+ and CD8+ cells were isolated from a separate portion of blood, and EM, CM, EMRA, SCM memory subpopulations were FACS sorted on days 30, 45 and {\ch 85}. For each sample we sequenced TCRalpha and TCRbeta repertoires. For both donors pre-infection PBMC repertoires were sampled in 2018 and 2019 for other projects. {\bf b,c, PCA of clonal temporal trajectories identifies three groups of clones with distinctive behaviours.} 
Left: first two principal components of the 1000 most abundant TCRbeta clonotype frequencies {\ch normalized by maximum value for each clonotype} in PBMC at post-infection timepoints. {\ch Color indicates hierarchical clustering results of principal components; symbol indicates if clonotype was called as significantly contracted from day 15 to day 85 (triangles), or expanded from day 15 to day 37 (circles) by both edgeR and NoisET (Fig. S5 shows overlap between clonal trajectory clusters and edgeR/NoisET hits).} Right: {\ch each curve shows the average $\pm$ 2.96 SE of normalized clonal frequencies from each cluster.} 
{\bf Contracting (d) and expanding (e) clones include both CD4+ and CD8+ T cells, and are less abundant in pre-infection repertoires.} T cell clones significantly contracted from day 15 to day {\ch 85} {\bf(d)} and significantly expanded from day 15 to day 37 {\bf(e)} were identified in both donors. The fraction of contracting {\bf(d)} and expanding {\bf(e)} TCRbeta clonotypes in the total repertoire ({\ch calculated as the sum of frequencies of these clonotypes in the second PBMC replicate at a given timepoint} and corresponding to the fraction of responding cells of all T cells) is plotted in log-scale for all reactive clones (left), reactive clones with the CD4 (middle) and  CD8 (right) phenotypes. Similar dynamics were observed in TCRalpha repertoires (Fig. S3), and for significantly expanded/contracted clones identified with the NoisET Bayesian differential expansion statistical model {\ch alone} (Fig. S4).
}

\label{fig1}
\end{figure*}

In the middle of March (day 0) donor W female and donor M (male, both  healthy young adults), returned to their home country from the one of the centers of the COVID-19 outbreak in Europe at the time. Upon arrival, according to local regulations, they were put into strict self-quarantine for 14 days. On day 3 of self-isolation both developed low grade fever, fatigue and myalgia, which lasted 4 days and was followed by a temporary loss of smell for donor M. On days 15, 30, 37, 45 and 85 we collected peripheral blood samples from both donors (Fig. 1a). The presence of IgG and IgM SARS-CoV-2 specific antibodies in the plasma was measured at all timepoints using SARS-CoV-2 S-RBD domain specific ELISA (Fig. S1). From each blood sample we isolated PBMCs (peripheral blood mononuclear cells, in two biological replicates), CD4+, and CD8+ T cells. Additionally, on days 30, 45 and {\ch 85} we isolated four T cell memory subpopulations (Fig. S2): Effector Memory (EM: CCR7-CD45RA-), Effector Memory with CD45RA re-expression (EMRA: CCR7-CD45RA+), Central Memory (CM: CCR7+CD45RA-), and Stem Cell-like Memory (SCM: CCR7+CD45RA+CD95+). From all samples we isolated RNA and performed TCRalpha and TCRbeta repertoire sequencing as previously described \cite{pogorelyy_persisting_2017}. For both donors, TCRalpha and TCRbeta repertoires were obtained for other projects one and two years prior to infection.  Additionally, TCR repertoires of multiple samples for donor M --  including sorted memory subpopulations -- are available from a published longitudinal TCR sequencing study after yellow fever vaccination (donor M1 samples in \cite{minervina_primary_2020}).

\subsection*{Two waves of T-cell clone response}

From previously described activated T cell dynamics for SARS-CoV-2 \cite{thevarajan_breadth_2020}, and immunization with live vaccines \cite{miller_human_2008}, the peak of the T cell expansion is expected around day 15 post-infection, and responding T cells significantly contract {\ch afterwards}. However, Weiskopf et al. \cite{weiskopf_phenotype_2020} reports an increase of SARS-CoV-2-reactive T cells at later timepoints, peaking in some donors after 30 days following symptom onset. To identify groups of T cell clones with similar dynamics in an unbiased way, we used Principal Component Analysis (PCA) in the space of T cell clonal trajectories (Fig. 1b and c). {\ch This exploratory data analysis method allows us to visualize major trends in the dynamics of abundant TCR clonotypes (occuring within top 1000 on any post-infection timepoints)  between multiple timepoints.}

In both donors, and in both TCRalpha and TCRbeta repertoires, we identified three clusters of clones with distinct dynamics. The first cluster (Fig. 1bc, purple) corresponded to abundant TCR clonotypes which had constant concentrations across timepoints, the second cluster (Fig. 1bc, green) showed contraction dynamics from day 15 to day {\ch 85}, and the third cluster (Fig. 1bc, yellow), showed an unexpected clonal expansion from day 15 with a peak on day 37 followed by contraction. The clustering and dynamics are similar in both donors and are reproduced in TCRbeta (Fig. 1bc) and TCRalpha (Fig. S3ab) repertoires. We next used edgeR, a software for differential gene expression analysis \cite{robinson_edger:_2010} {\ch and NoisET, a Bayesian differential expansion model \cite{puelma_touzel_inferring_2020}}, to specifically detect changes in clonotype concentration between pairs of timepoints in a statistically reliable way {\ch and without limiting the analysis to the most abundant clonotypes}. {\ch Both NoisET and} edgeR use biological replicate samples collected at each timepoint to train a noise model for sequence counts. {\ch Results for the two models were similar (Fig. S4) and we conservatively defined as expanded or contracted the clonotypes that were called by both models simultaneously.} We identified {\ch 291} TCRalpha and {\ch 295} TCRbeta clonotypes in donor W, and {\ch 607} TCRalpha and {\ch 616} TCRbeta in donor M significantly contracted from day 15 to day {\ch 85} (largely overlapping with cluster 2 of clonal trajectories, Fig. S5). {\ch 176} TCRalpha and {\ch 278} TCRbeta for donor W, and {\ch 293} TCRalpha and {\ch 427} TCRbeta clonotypes for donor M were significantly expanded from day 15 to 37 (corresponding to cluster 3 of clonal trajectories). 

Note that, to identify putatively SARS-CoV-2 reactive clones, we only used  post-infection timepoints, so that our analysis can be reproduced in other patients and studies where pre-infection timepoints are unavailable. However, tracking the identified responding clones back to pre-infection timepoints reveals strong clonal expansions from pre- to post-infection (Fig. 1de, Fig. S3cd). For brevity, we further refer to clonotypes significantly contracted from day 15 to {\ch 85} as \textit{contracting} clones and clonotypes significantly expanding from day 15 to 37 as \textit{expanding} clones. {\ch Contracting clones corresponded to 2.5\% and 0.9\% of T cells on day 15 post-infection, expanding clones reached 1.1\% and 0.8\% on day 37 for donors M and W respectively (Fig. 1de, left). This magnitude of the T cell response is of the same order of magnitude as previously observed after live yellow fever vaccine immunization of donor M (6.7\%  T cells on day 15 post-vaccination).} For each contracting and expanding clone we determined their CD4/CD8 phenotype using separately sequenced repertoires of CD4+ and CD8+ subpopulations (see Methods). Both CD4+ and CD8+ subsets participated actively in the response (Fig. 1de). Interestingly, clonotypes expanding after day 15 were significantly biased towards the CD4+ phenotype, while contracting clones had balanced CD4/CD8 phenotype fractions in both donors (Fisher exact test, $p<0.01$ for both donors). 

\begin{figure*}
\noindent\includegraphics[width=\ff\linewidth]{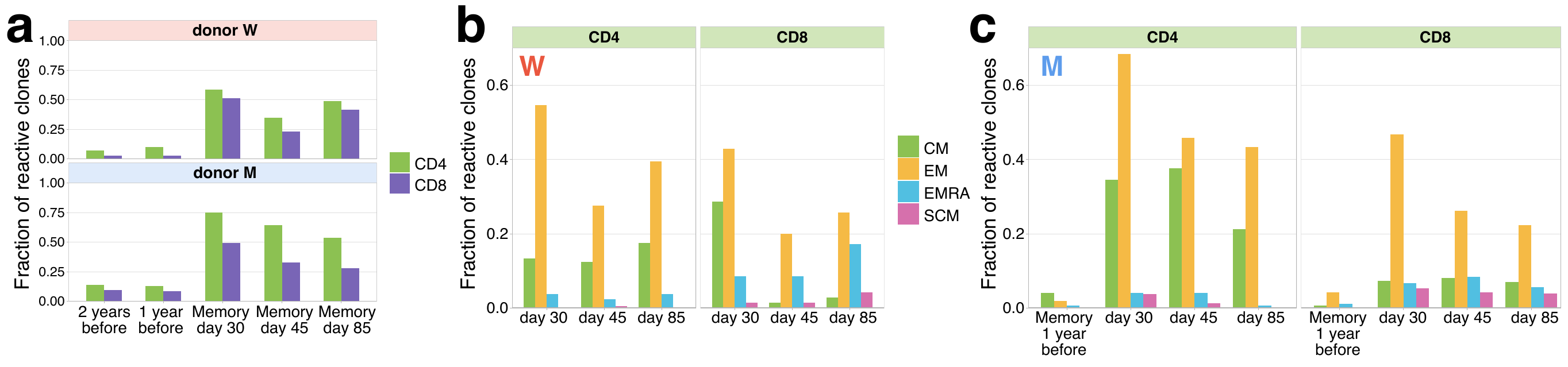}
\caption{{\bf Memory phenotypes of responding clonotypes contracting after day 15. a, A large fraction of contracting clonotypes is identified in T cell memory subsets after infection.} Bars show the fraction of contracting CD4+ and CD8+ TCRbeta clonotypes present in 2-year; 1-year pre-infection PBMC; in at least one of memory subpopulation sampled on day 30, day 37 and day 85 post infection. {\bf b,c Responding clones are found in different memory subsets.} {\ch Fraction of CD4+ (left panels) and CD8+ (right panels) contracting clones of donor W {\bf(b)} and M {\bf(c)} that were identified in each memory subpopulation repertoire at each timepoint. For both donors, CD4+ clonotypes were found predominantly in Central Memory (CM) and  Effector Memory (EM), while CD8+ T cells were enriched in EMRA compartment.}{\bf c,} For donor M, CD4+ contracting clonotypes are also identified in memory subsets 1 year before the infection, with a bias towards the CM subpopulation {\ch and a group of CD8+ clones is found in the pre-infection EM subpopulation.} 
}
\label{fig2}
\end{figure*}

\subsection*{Memory formation and pre-existing memory}

On days 30, 45 {\ch and 85} we identified both contracting (Fig. 2a-c) and expanding (Fig. S6a-c) T cell clones in the memory subpopulations of peripheral blood. Both CD4+ and CD8+ responding clones were found in the CM and EM subsets, however CD4+ were more biased towards CM ({\ch with exception of donor W day 30 timepoint, where a considerable fraction of CD8+ clones were found in CM}), and CD8+ clones more represented in the EMRA subset. A small number of both CD4+ and CD8+ responding clonotypes were also identified in the SCM subpopulation, which was previously shown to be a long-lived T cell memory subset \cite{fuertes_marraco_long-lasting_2015}. {\ch Note that we sequenced more cells from PBMC than from the memory subpopulations (Table S1), so that some low-abundant responding T cell clones are not sampled in the memory subpopulations.}
Intriguingly, a number of responding CD4+ clones, and {\ch fewer} CD8+ clones, were also represented in the repertoires of both donors 1 and 2 years before the infection. Pre-existing clones were expanded after infection, and contracted afterwards for both donors (Fig. S7). For donor M, for whom we had previously sequenced memory subpopulations before the infection \cite{minervina_primary_2020}, we were able to identify pre-existing SARS-CoV-2-reactive CD4+ clones in the CM subpopulation 1 year before the infection {\ch and a group of CD8+ clones in the pre-infection EM subpopulation}. Interestingly, on day 30 after infection the majority of pre-infection CM clones were detected in the EM subpopulation, suggesting recent T cell activation and a switch of the phenotype from memory to effector. These clones might represent memory T cells cross-reactive for other infections, e.g. other human coronaviruses.

A search for TCRbeta amino acid sequences of responding clones in VDJdb \cite{bagaev_vdjdb_2020} --- a database of TCRs with known specificities --- resulted in essentially no overlap with {\ch TCRs not specific for SARS-CoV-2 epitopes: only two clonotypes matched. One match corresponded to the CMV (cytomegalovirus) epitope presented by the HLA-A*03 MHC allele, which is absent in both donors (Table S2), and a second match was for Influenza A virus epitope presented by HLA-A*02 allele. The absence of matches suggests that contracting and expanding clones are unlikely to be specific for immunodominant epitopes of common pathogens covered in VDJdb. We next asked if we could map specificites of our responding clones to SARS-CoV-2 epitopes.}

\subsection*{Validation by MHC tetramer-staining assay}

{\ch On day 25 post-infection donor M participated in study by Shomuradova et al. \cite{Shomuradova2020} (as donor p1434), where his CD8+ T cells were stained with HLA-A*02:01-YLQPRTFLL MHC-I tetramer. TCRalpha and TCRbeta of FACS-sorted tetramer-positive cells were sequenced and deposited to VDJdb (see \cite{Shomuradova2020} for the experimental details). We matched these tetramer-specific TCR sequences to our longitudinal dataset (Fig.~3a for TCRbeta and Fig.~S8 for TCRalpha). We found that their frequencies were very low on pre-infection timepoints and monotonically decreased from their peak on day 15 ($7.1\cdot10^{-4}$ fraction of bulk  TCRbeta repertoire) to day 85 ($1.3\cdot10^{-5}$ fraction), in close analogy to our contracting clone set.
Among the tetramer positive clones that were abundant on day 15 (with bulk frequency $>10^{-5}$), 17 out of 18 or TCRbetas and 12 out of 15 TCRalphas were independently identified as contracting by our method.
  
}

\begin{figure*}
\noindent\includegraphics[width=\ff\linewidth]{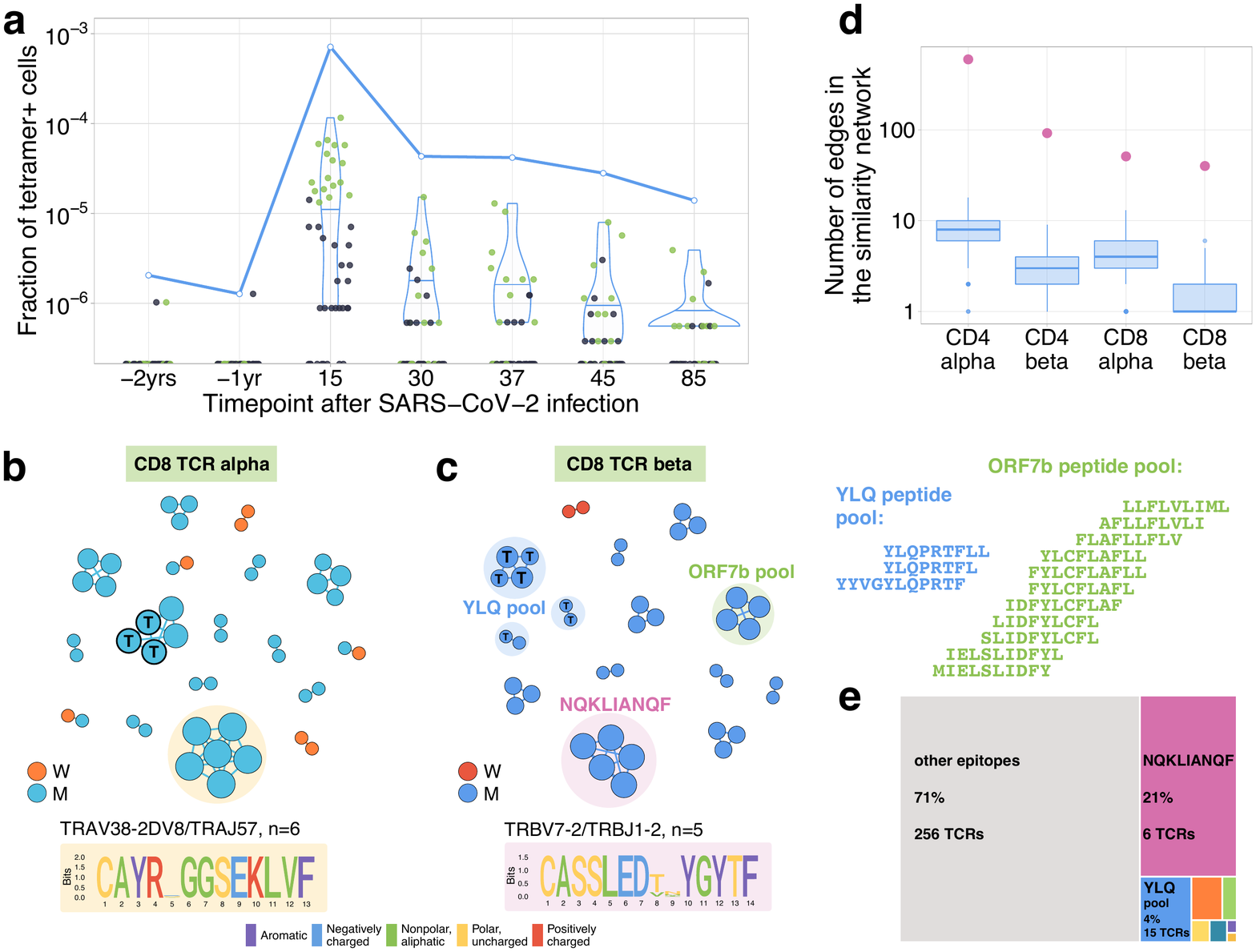}
\caption{
{\ch {\bf a, SARS-CoV-2-specific TCRs are independently identified by clonal contraction.} Each dot corresponds to the frequency of HLA-A*02:01-YLQPRTFLL-tetramer specific TCRbeta clonotype in bulk repertoire of donor M (donor p1434 from \cite{Shomuradova2020}) at each timepoint.
    Green dots correspond to clonotypes independently identified as contracting in our longitudinal dataset. Blue line shows the cumulative frequency of tetramer specific TCRbeta clonotypes.   {\bf b, c Analysis of TCR amino acid sequences of contracting CD8+ clones reveals distinctive motifs.} For each set of CD8alpha, and CD8beta contracted clonotypes, we constructed a separate similarity network. Each vertex in the similarity network corresponds to a contracting clonotype. An edge indicates 2 or less amino acid mismatches in the CDR3 region, and identical V and J segments. Networks are plotted separately for CD8alpha {\bf(b)} and CD8beta {\bf(c)} contracting clonotypes. Clonotypes without neighbours are not shown. Sequence logos corresponding to the largest clusters are shown under the corresponding network plots. `T' on vertices indicate TCRbeta clonotypes confirmed by HLA-A*02:01-YLQPRTFLL tetramer staining. Shaded colored circles ({\bf(c)}) indicate clonotypes with large number of matches to CD8+ TCRs recognising SARS-CoV-2 peptides pools from ref. \cite{snyder_magnitude_2020} (MIRA peptide dataset). Lists of peptides in YLQ and ORF7b peptide pools are shown on the right. {\bf d, Sequence convergence among contracting clonotypes.} The number of edges in each group is shown by pink dots and is compared to the distribution of that number in 1000 random samples of the same size from the relevant repertoires at day 15 (blue boxplots). {\bf e, } Fraction of TCRbeta clonotypes with matches in the MIRA dataset (coloured rectangles) out of all responding CD8+ TCRbeta clonotypes in donor M on day 15.
  } 
  }
\label{fig3}
\end{figure*}

\subsection*{TCR sequence motifs of responding clones}

It was previously shown that TCRs recognising the same antigens frequently have highly similar TCR sequences \cite{dash_quantifiable_2017,glanville_identifying_2017}. To identify motifs in TCR amino acid sequences, we plotted similarity networks for significantly contracted (Fig. 3bc, Fig. 4ab) and expanded (Fig. S9b-e) clonotypes. {\ch The number of edges in all similarity networks except CD8+ expanding clones was significantly larger than would expected by randomly sampling the same number of clonotypes from the corresponding repertoire (Fig. 3d and Fig. S9a).} In both donors we found clusters of highly similar clones in both CD4+ and CD8+ subsets for expanding and contracting clonotypes. Clusters were largely donor-specific, as expected, since our donors have dissimilar HLA alleles (SI Table 1) and thus each is likely to present a non-overlapping set of T cell antigens. The largest cluster, described by the motif TRAV35-CAGXNYGGSQGNLIF-TRAJ42, was identified in donor M's CD4+ contracting alpha chains. Clones from this cluster constituted 16.3\% of all of donor M's CD4+ responding cells on day 15, suggesting a response to an immunodominant CD4+ epitope in the SARS-CoV-2 proteome. The high similarity of the TCR sequences of responding clones in this cluster allowed us to independently identify motifs from donor M's CD4 alpha contracting clones using the ALICE algorithm \cite{pogorelyy_detecting_2019} (Fig. S10). While the time dependent methods (Fig. 1) identify abundant clones, the ALICE approach is complementary to both edgeR and NoisET as it identifies clusters of T cells with similar sequences independently of their individual abundances. 

\begin{figure*}
\noindent\includegraphics[width=\ff\linewidth]{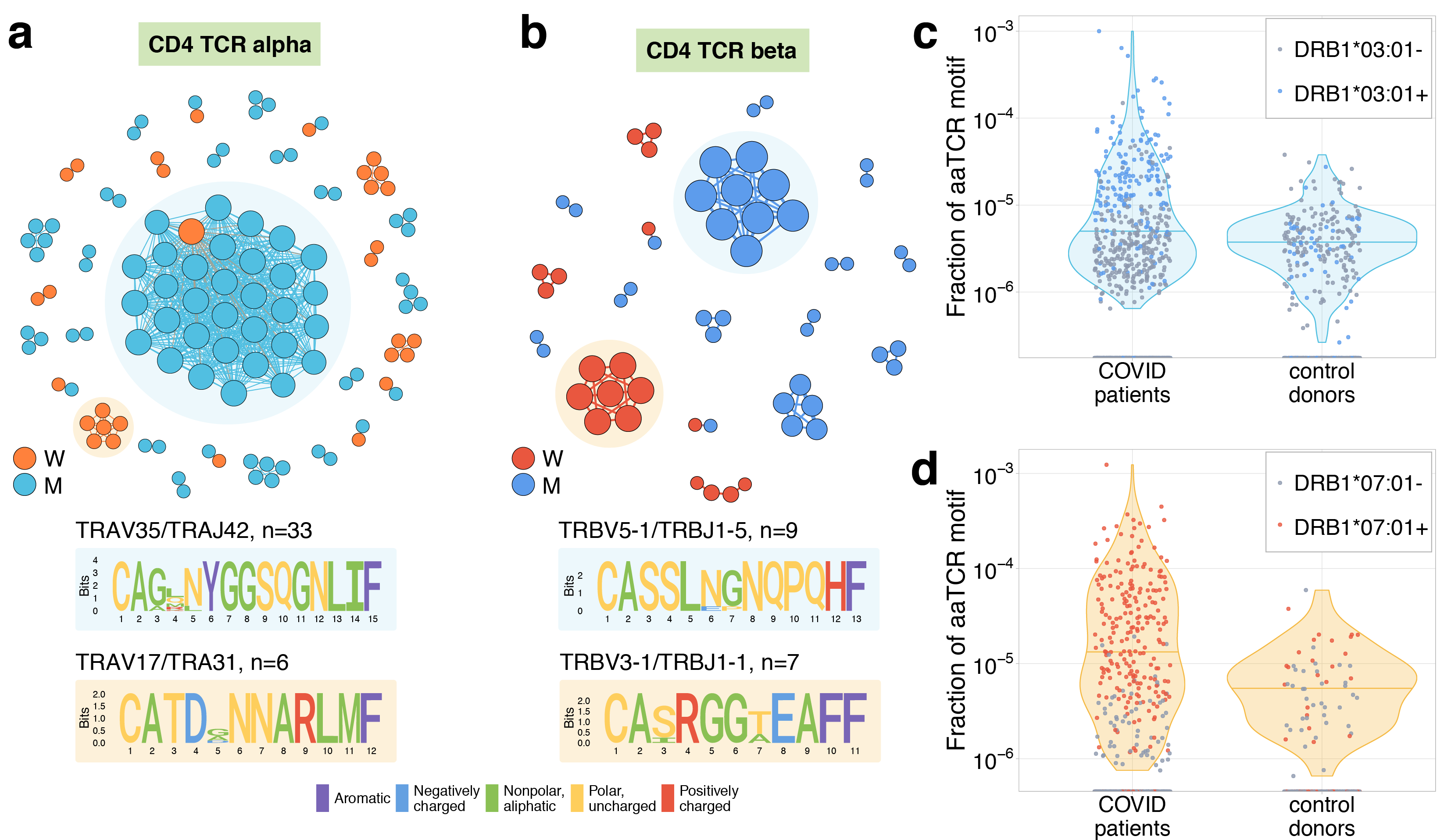}
\caption{{\bf a, Analysis of TCR amino acid sequences of CD4+ contracting clones reveal distinctive motifs.} Each vertex in the similarity network corresponds to a contracting clonotype. An edge indicates 2 or less amino acid mismatches in the CDR3 region (and identical V and J segments). Networks are plotted separately for CD4alpha {\bf(a)}, CD4beta {\bf(b)}, contracting clonotypes. Clonotypes without neighbours are not shown. Sequence logos corresponding to the largest clusters are shown under the corresponding network plots. {\ch {\bf c, d,} Clonotypes forming the two largest motifs are significantly more clonally expanded (p$<$0.001, one sided t-test) in a cohort of COVID-19 patients \cite{snyder_magnitude_2020} than in a cohort of control donors \cite{emerson_immunosequencing_2017}. Each dot corresponds to the total frequency of clonotypes from motifs shaded on {\bf (b)} in the TCRbeta repertoire of a given donor. Colored dots show donors predicted to share HLA-DRB1*07:01 allele with donor W {\bf(c)}, or HLA-DRB1*03:01-DQB1*02:01 haplotype with donor M {\bf(d)}. }
}
\label{fig4}
\end{figure*}

\subsection*{Mapping TCR motifs to SARS-CoV-2 epitopes}

{\ch In CD8+ T cells, 3 clusters of highly similar TCRbeta clonotypes in donor M and one cluster of TCRalpha clonotypes correspond to YLQPRTFLL-tetramer-specific TCR sequences described above. To map additional specificities for CD8+ TCRbetas, we used a large set of SARS-CoV-2-peptide specific TCRbeta sequences from \cite{snyder_magnitude_2020} obtained using Multiplex Identification of Antigen-specific T cell Receptors Assay (MIRA) with combinatorial peptide pools \cite{klinger_multiplex_2015}. For each responding CD8+ TCRbeta we searched for the identical or highly similar (same VJ combination, up to one mismatch in CDR3aa) TCRbeta sequences specific for given SARS-CoV-2 peptides. A TCRbeta sequence from our set was considered mapped to a given peptide if it had at least two highly similar TCRbeta sequences specific for this peptide in the MIRA experiment. This procedure yielded unambiguous matches for 32 CD8+ TCRbetas --- just one clonotype was paired to two peptide pools (Table S3).
The vast majority of matches to MIRA corresponded to groups of contracting clones. As expected, we found that all clusters corresponding to HLA-A*02:01-YLQPRTFLL MHC-I tetramer-specific TCRs were matched to the peptide pool YLQPRTFL,YLQPRTFLL,YYVGYLQPRTF in the MIRA dataset. Another large group of matches corresponded to the HLA-B*15:01-restricted \cite{stamatakis_generation_2020} NQKLIANQF epitope. Interestingly, clonotypes corresponging to this cluster together made up 21\% of the CD8+ immune response on day 15, suggesting immunodominance of this epitope. Two TCRbeta clonotypes mapped to this epitope were identified in Effector Memory subset one year before the infection, suggesting potential cross-reactive response. We speculate, that this response might be initially triggered by NQKLIANAF, a homologous HLA-B*15:01 epitope from HKU1 or OC43, common human betacoronaviruses. To predict potential pairings between TCRalpha and TCRbeta motifs, we used a method of alpha/beta clonal trajectory matching described in \cite{minervina_primary_2020} (see Methods for details). We found consistent pairing between one of the motifs in TCRalpha to the largest motif in TCRbeta T cells, which is associated to HLA-B*15:01-NQKLIANQF. 

\subsection*{Validation of CD4+ COVID-19 HLA-restricted specificity by cohort association analysis}

At the time of writing, no data on TCR sequences specific to MHC-II class epitopes exist to map specificities of CD4+ T-cells in a similar way as we did with MIRA-specific TCRs. However, a recently published database of 1414 bulk TCRbeta repertoires from COVID-19 patients allowed us to confirm the SARS-CoV-2 specificity of contracting clones indirectly. Public TCRbeta sequences that can recognize SARS-CoV-2 epitopes are expected to be clonally expanded and thus sampled more frequently in the repertoires of COVID-19 patients than in control donors. In Fig. 4cd we show that the total frequency of TCRbeta sequences forming the largest cluster in donor M (Fig. 4c) and donor W (Fig. 4d) is significantly larger in the COVID-19 cohort than in the healthy donor cohort from ref. \cite{emerson_immunosequencing_2017}, suggesting antigen-dependent clonal expansion. We hypothesized that the difference between control and COVID-19 donors in motif abundance should be even larger if we restrict the analysis to donors sharing the HLA allele that presents the epitope.
Unfortunately, HLA-typing information is not yet available for the COVID-19 cohort. However, using sets of HLA-associated TCRbeta sequences from ref. \cite{Dewitt2018}, we could build a simple classifier to predict the HLA alleles of donors from both the control and COVID-19 cohorts exploiting the presence of TCRbeta sequences associated with certain HLA alleles (see Methods for details). 
We found that the CD4+ TCRbeta motif from donor W occurs preferentially in donors predicted to have DRB1*07:01 allele, while the motif from donor M appears to be associated with  HLA-DRB1*03:01-DQB1*02:01 haplotype. The frequency of sequences corresponding to these motifs can then be used to identify SARS-CoV-2 infected donors with matching HLA alleles (Fig. S11).
}

\section*{Discussion}
Using longitudinal repertoire sequencing, we identified a group of CD4+ and CD8+ T {\ch cell} clones that contract after recovery from a SARS-CoV-2 infection. Our response timelines agree with T cell dynamics reported by Theravajan et al. \cite{thevarajan_breadth_2020} for mild COVID-19, as well as with dynamics of T cell response to live vaccines \cite{miller_human_2008}. {\ch We further mapped the specificities of contracting CD8+ T cells using sequences of SARS-CoV-2 specific T cells identified with tetramer staining in the same donor, and as well as the large set of SARS-CoV-2 peptide stimulated TCRbeta sequences from ref. \cite{snyder_magnitude_2020}. For large CD4+ TCRbeta motifs we show strong association with COVID-19 by analysing the occurence patterns and frequencies of these sequences in a large cohort of COVID-19 patients.}
Surprisingly, in both donors we also identified a group of predominantly CD4+ clonotypes which expanded from day 15 to day 37 after the infection. One possible explanation for this second wave of expansion is the priming of CD4+ T cells by antigen-specific B-cells, but there might be other mechanisms such as the migration of SARS-CoV-2 specific T cells from lymphoid organs {\ch or bystander activation of non-SARS-CoV-2 specific T cells}. It is also possible that later expanding T cells are triggered by another infection, simultaneously and asymptomatically occurring in both donors around day 30. {\ch In contrast with the first wave of response identified by contracting clones, for now we do not have confirmation that this second wave of expansion corresponds to SARS-CoV-2 specific T cells. Accumulation of TCR sequences for CD4+ SARS-CoV-2 epitope specific T cells may further address this question.}
We showed that a large fraction of putatively SARS-CoV-2 reactive T cell clones are later found in memory subpopulations {\ch and remain there at least 3 months after infection. Importantly, some of responding clones are found in long-lived stem cell-like (SCM) memory subset, as also reported for SARS-CoV-2 convalescent patients in ref. \cite{sekine_robust_2020}. } 
A subset of CD4+ clones were identified in pre-infection central memory subsets, {\ch and a subset of CD8+ T cells were found in effector memory. Among these are CD8+ clones recognising NQKLIANQF, an immunodominant HLA-B*15:01 restricted SARS-CoV-2 epitope, for which homologous epitope differing by 1 aa mismatch exists in common human betacoronaviruses. 
} The presence of SARS-CoV-2 cross-reactive CD4+ T cells in healthy individuals was recently demonstrated \cite{braun_presence_2020,grifoni_targets_nodate,le_bert_sars-cov-2-specific_2020,meckiff_single-cell_2020,bacher_pre-existing_2020,peng_broad_2020}. Our data further suggests that cross-reactive CD4+ and CD8+ T cells can participate in the response \textit{in vivo}. 
It is interesting to ask if the presence of cross-reactive T cells before infection is linked to the mildness of the disease {\ch(with predicted HLA-B*15:01 cross-reactive epitope described above as a good starting point)}. Larger studies with cohorts of severe and mild cases with pre-infection timepoints are needed to address this question.

\section*{Methods}
\subsection*{Donors and blood samples}
Peripheral blood samples from two young healthy adult volunteers, donor W (female) and donor M (male) were collected with written informed consent in a certified diagnostics laboratory. Both donors gave written informed consent to participate in the study under the declaration of Helsinki. HLA alleles of both donors (Table S2) were determined by an in-house cDNA high-throughput sequencing method.

\subsection*{SARS-CoV-2 S-RBD domain specific ELISA}
An ELISA assay kit developed by the National Research Centre for Hematology was used for detection of anti-S-RBD IgG according to the manufacturer's protocol. The relative IgG level (OD/CO) was calculated by dividing the OD (optical density) values by the mean OD value of the cut-off positive control serum supplied with the Kit (CO). OD values of d37, d45 and d85 samples for donor M exceeded the limit of linearity for the Kit. In order to properly compare the relative IgG levels between d30, d37, d45 and d85, these samples were diluted 1:400 instead of 1:100, the ratios d37:d30 and d45:d30 and d85:d30 were calculated and used to calculate the relative IgG level of d37, d45 and d85 by multiplying d30 OD/CO value by the corresponding ratio. {\ch Relative anti-S-RBD IgM level was calculated using the same protocol with anti-human IgM-HRP conjugated secondary antibody. Since the control cut-off serum for IgM was not available from the Kit, on Fig. S1b. we show OD values for nine biobanked pre-pandemic serum samples from healthy donors.}

\subsection*{Isolation of PBMCs and T cell subpopulations}
PBMCs were isolated with the Ficoll-Paque density gradient centrifugation protocol. CD4+ and CD8+ T cells were isolated from PBMCs with Dynabeads CD4+ and CD8+ positive selection kits (Invitrogen) respectively. For isolation of EM, EMRA, CM and SCM memory subpopulations we stained PBMCs with the following antibody mix: anti-CD3-FITC (UCHT1, eBioscience), anti-CD45RA-eFluor450 (HI100, eBioscience), anti-CCR7-APC (3D12, eBioscience), anti-CD95-PE (DX2, eBioscience). Cell sorting was performed on FACS Aria III, all four isolated subpopulations were lysed with Trizol reagent immediately after sorting. 

\subsection*{TCR library preparation and sequencing}
TCRalpha and TCRbeta cDNA libraries preparation was performed as previously described in \cite{pogorelyy_persisting_2017}. RNA was isolated from each sample using Trizol reagent according to the manufacturer's instructions. A universal primer binding site, sample barcode and unique molecular identifier (UMI) sequences were introduced using the 5'RACE technology with TCRalpha and TCRbeta constant segment specific primers for cDNA synthesis. cDNA libraries were amplified in two PCR steps, with introduction of the second sample barcode and Illumina TruSeq adapter sequences at the second PCR step. 
Libraries were sequenced using the Illumina NovaSeq platform (2x150bp read length).

\subsection*{TCR repertoire data analysis}

{\bf Raw data preprocessing.} 
Raw sequencing data was demultiplexed and UMI guided consensuses were built using migec v.1.2.7 \cite{shugay_towards_2014}. Resulting UMI consensuses were aligned to V and J genomic templates of the TRA and TRB locus and assembled into clonotypes with mixcr v.2.1.11 \cite{bolotin_mixcr:_2015}. See Table S1 for the number of cells, UMIs and unique clonotypes for each sample.

{\bf Identification of clonotypes with active dynamics.} Principal component analysis (PCA) of clonal trajectories was performed as described before \cite{minervina_primary_2020}. First we selected clones which were present among the top 1000 abundant in any of post-infection PBMC repertoires, {\ch including biological replicates, i.e. considered clone abundant if it was found within top 1000 most abundant clonotypes in at least one of the replicate samples at one timepoint.} Next, for each such abundant clone we calculated its frequency at each post-infection timepoint and divided this frequency by the maximum frequency of this clone for normalization. Then we performed PCA on the resulting normalized clonal trajectory matrix and identified three clusters of trajectories with hierarchical clustering with {\ch average} linkage, using Euclidean distances between trajectories.
 We identify statistically significant contractions and expansions with edgeR as previously described \cite{pogorelyy_precise_2018}, using FDR adjusted $p<0.01$ and $\log_2$ fold change threshold of 1. NoisET implements the Bayesian detection method described in \cite{puelma_touzel_inferring_2020}. Briefly, a two-step noise model accounting for cell sampling and expression noise is inferred from replicates, and a second model of expansion is learned from the two timepoints to be compared. The procedure outputs the posterior probability of expansion or contraction, and the median estimated $\log_2$ fold change, whose thresholds are set to 0.05 and 1 respectively.

{\ch {\bf Mapping of COVID-19 associated TCRs to the MIRA database.} 
TCRbeta sequences from T cells specific for SARS-CoV-2 peptide pools MIRA (ImmuneCODE release 2) were downloaded from \url{https://clients.adaptivebiotech.com/pub/covid-2020}. V and J genomic templates were aligned to TCR nucleotide sequences from the MIRA database using mixcr 2.1.11. We consider a TCRbeta from MIRA matched to a TCRbeta from our data, if it had the same V and  J and at most one mismatch in CDR3 amino acid sequence. We consider a TCRbeta sequence mapped to an epitope if it has at least two identical or highly similar (same V, J and up to one mismatch in CDR3 amino acid sequence) TCRbeta clonotypes reactive for this epitope in the MIRA database. 

{\bf Computational alpha/beta pairing by clonal trajectories.} 
Computational alpha/beta pairing was performed as described in \cite{minervina_primary_2020}. For each TCRbeta  we determine the TCRalpha with the closest clonal trajectory (Tables S3 and S5). We observe no stringent pairings between TCRbeta and TCRbeta motifs with exception of two contracting CD8 TCRbeta clusters: TRBV7-2/TRBJ1-2 NQKLIANQF-associated clones from donor M paired to TRAV21/TRAJ40 alphas from the same cluster (CASSLEDTNYGYTF-CAVHSSGTYKYIF and CASSLEDTIYGYTF-CAALTSGTYKYIF), and TRBV7-9/TRBJ2-3 beta cluster paired to largest alpha cluster (CASSPTGRGRTDTQYF-CAYRSGGSEKLVF and CASSPTGRGGTDTQYF-CAYRRPGGEKLTF).  

{\bf Computational prediction of HLA-types.}
To predict HLA-types from TCR repertoires of COVID-19 cohort we used sets of HLA-associated TCR sequences from \cite{Dewitt2018}. We use TCRbeta repertoires of 666 donors from cohort from \cite{emerson_immunosequencing_2017}, for which HLA-typing information is available in ref. \cite{Dewitt2018} as a training set to fit logistic regression model, where presence or absense of given HLA-allele is an outcome, and the number of allele-associated sequences in repertoire, as well as the total number of unique sequences in the repertoire, are the predictors. A separate logistic regression model was fitted for each set of HLA-associated sequences from ref. \cite{Dewitt2018}, and then used to predict the probability $p$ that a donor from the COVID-19 cohort has this allele. Donors with $p<0.2$ were considered negative for a given allele. 
}

\subsection*{Data availability}
Raw sequencing data are deposited to the Short Read Archive (SRA) accession: PRJNA633317.
Processed TCRalpha and TCRbeta repertoire datasets, resulting repertoires of SARS-CoV-2-reactive clones, and raw data preprocessing instructions can be accessed from: \url{https://github.com/pogorely/Minervina_COVID}.

\section*{Acknowledgments}
The study was supported by RSF 20-15-00351. E.R and A.F. are supported by the DFG Excellence Cluster Precision Medicine in Chronic Inflammation (Exc2167) and the DFG grant n. 4096610003, M.B.K., T.M., and A.M.W. are supported by European Research Council COG 724208. I.Z.M. is supported by RFBR 19-54-12011 and 18-29-09132. D.M.C. is supported by the grant from Ministry of Science and Higher Education of the Russian Federation 075-15-2019-1789.

\bibliographystyle{pnas}

\setcounter{figure}{0}
\setcounter{table}{0}
\renewcommand{\thefigure}{S\arabic{figure}}
\renewcommand{\thetable}{S\arabic{table}}

\begin{figure*}[p]
\noindent\includegraphics[width=0.8\linewidth]{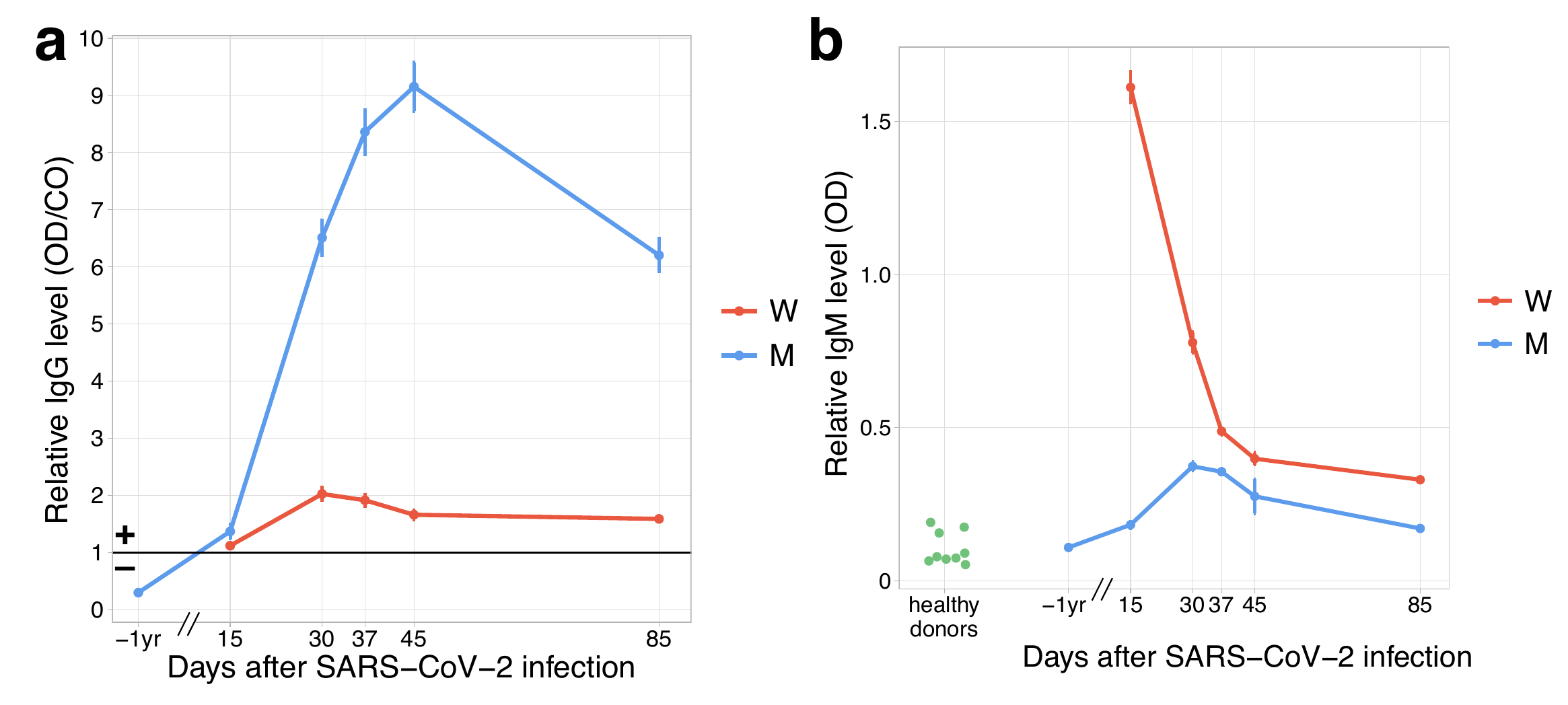}
\caption{{\bf Both donors developed anti-SARS-CoV-2 IgG and IgM responses by day 15 post infection. a,}  The relative level of SARS-CoV-2 S-RBD domain specific IgG  (y-axis) is plotted against time. Solid black line shows the threshold for positive testing. {\bf b,} Relative IgM levels in donors M and W are shown over time. Relative IgM levels for pre-pandemic serum samples from healthy donors are shown on the left (green dots). }
\end{figure*}

\begin{figure*}[p]
\noindent\includegraphics[width=\linewidth]{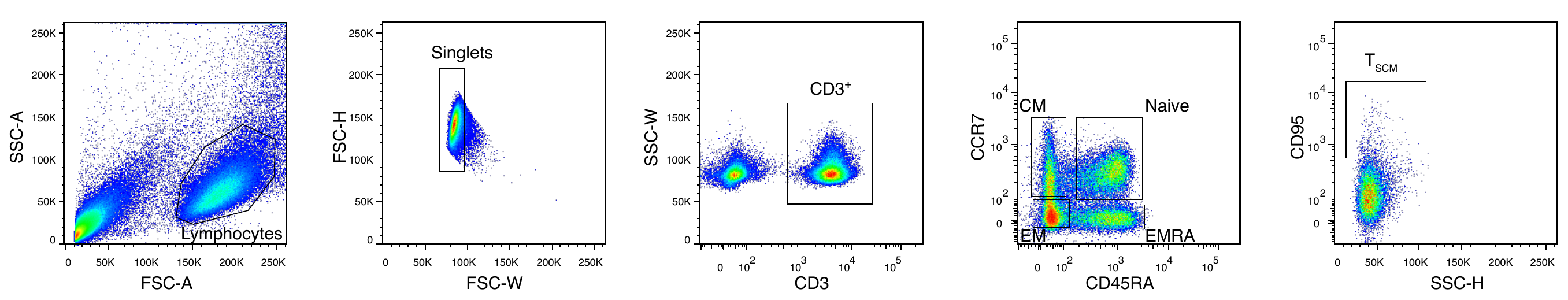}
\caption{ {\bf Memory subpopulation gating strategy.} Three populations of memory T cells: EM, CM and EMRA are defined by CCR7 and CD45RA markers, SCM are distinguished from naive CCR7+ CD45RA+ T cells by CD95 expression.
}
\end{figure*}

\begin{figure*}[p]
\noindent\includegraphics[width=\linewidth]{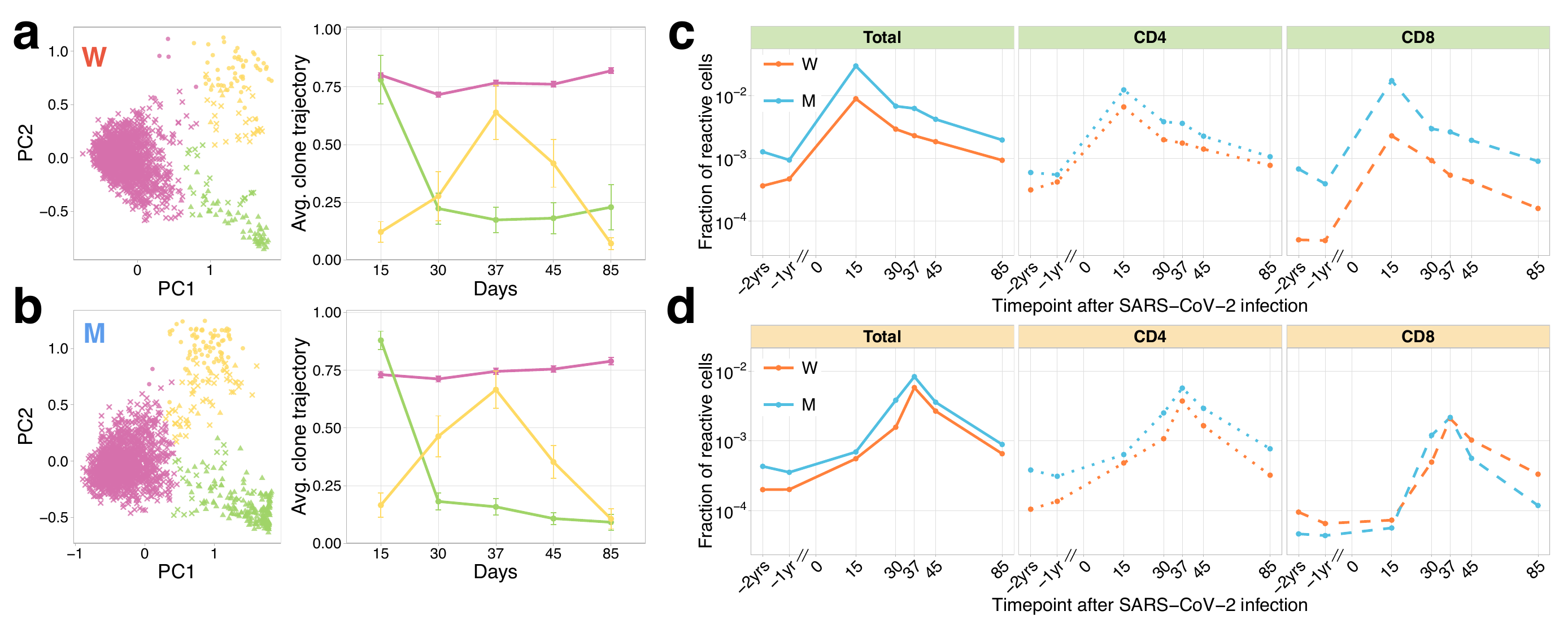}
\caption{{\bf Longitudinal tracking of T cell clones after mild COVID-19 with TCRalpha repertoire sequencing. a,b, PCA of clonal temporal trajectories identifies three groups of clones with distinctive behaviours.} 
Left: first two principal components of the 1000 most abundant TCRalpha clonotype frequencies {\ch normalized by maximum value for each clonotype} in PBMC at post-infection timepoints. {\ch Color indicates hierarchical clustering results of principal components; symbol indicates if clonotype was called as significantly contracted from day 15 to day 85 (triangles), or expanded from day 15 to day 37 (circles) by both edgeR and NoisET (Fig. S5 shows overlap between clonal trajectory clusters and edgeR/NoisET hits).} Right: {\ch each curve shows the average $\pm$ 2.96 SE of normalized clonal frequencies from each cluster.} 
{\bf Contracting (c) and expanding (d) clones include both CD4+ and CD8+ T cells, and are less abundant in pre-infection repertoires.} T cell clones significantly contracted from day 15 to day {\ch 85} {\bf(c)} and significantly expanded from day 15 to day 37 {\bf(d)} were identified in both donors. The fraction of contracting {\bf(c)} and expanding {\bf(d)} TCRalpha clonotypes in the total repertoire ({\ch calculated as the sum of frequencies of these clonotypes in the second PBMC replicate at a given timepoint} and corresponding to the fraction of responding cells of all T cells) is plotted in log-scale for all reactive clones (left), reactive clones with the CD4 (middle) and  CD8 (right) phenotypes. 
}
\end{figure*}

\begin{figure*}[p]
\noindent\includegraphics[width=0.8\linewidth]{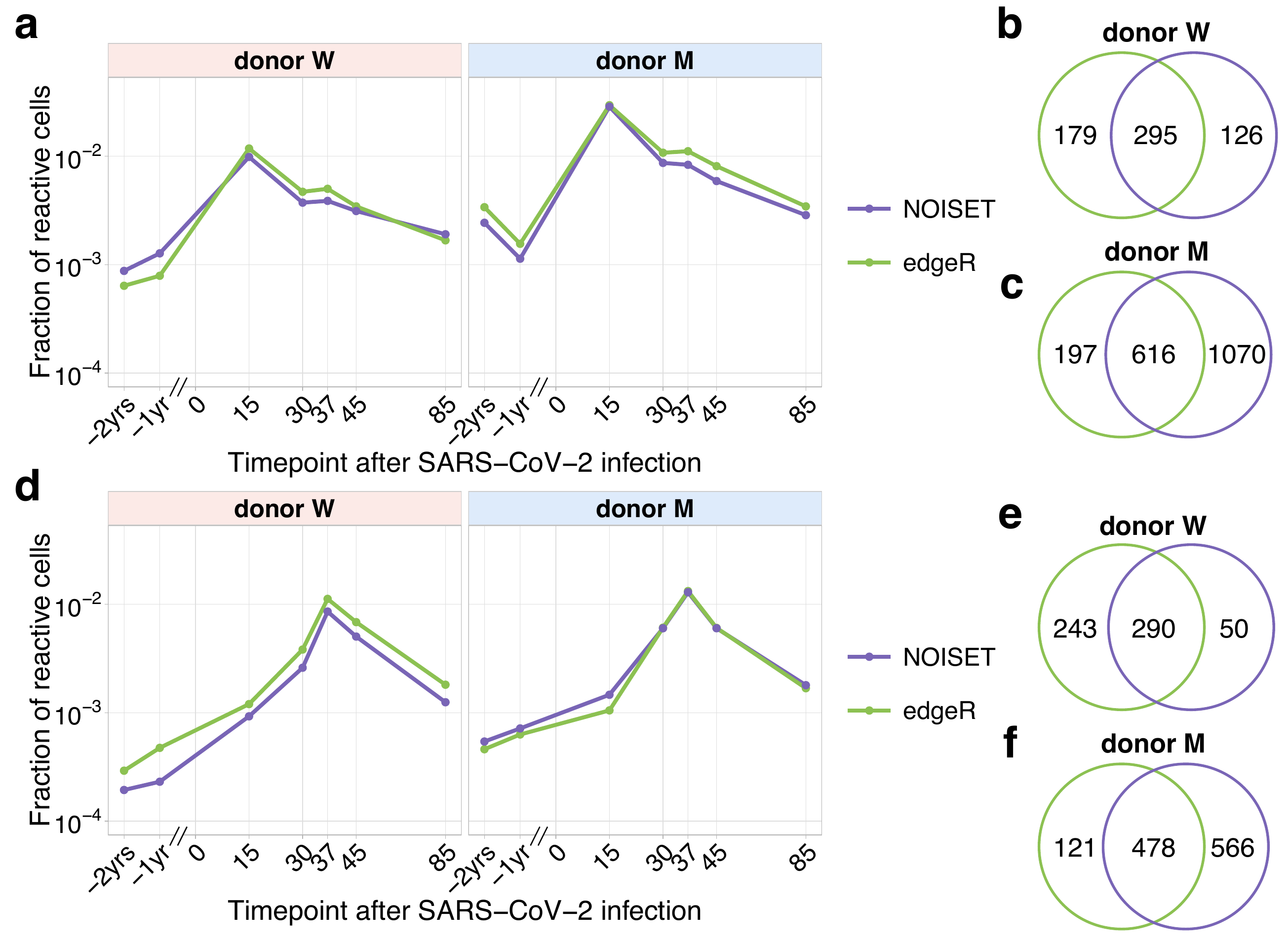}
\caption{ {\bf Comparison of edgeR and NoisET clonal expansion detection procedures.} The fraction (plotted in the log-scale) of contracting {\bf(a)} and expanding {\bf(d)} TCRbeta clonotypes in the total repertoire was estimated using subsets of expanded and contracted clones called by edgeR (green) and NoisET (purple) models. Overlaps for contracted clones {\bf(b,c)} and expanded clones {\bf(e,f)} identified by both models are shown on the right.
}
\end{figure*}

\begin{figure*}[p]
\noindent\includegraphics[width=0.9\linewidth]{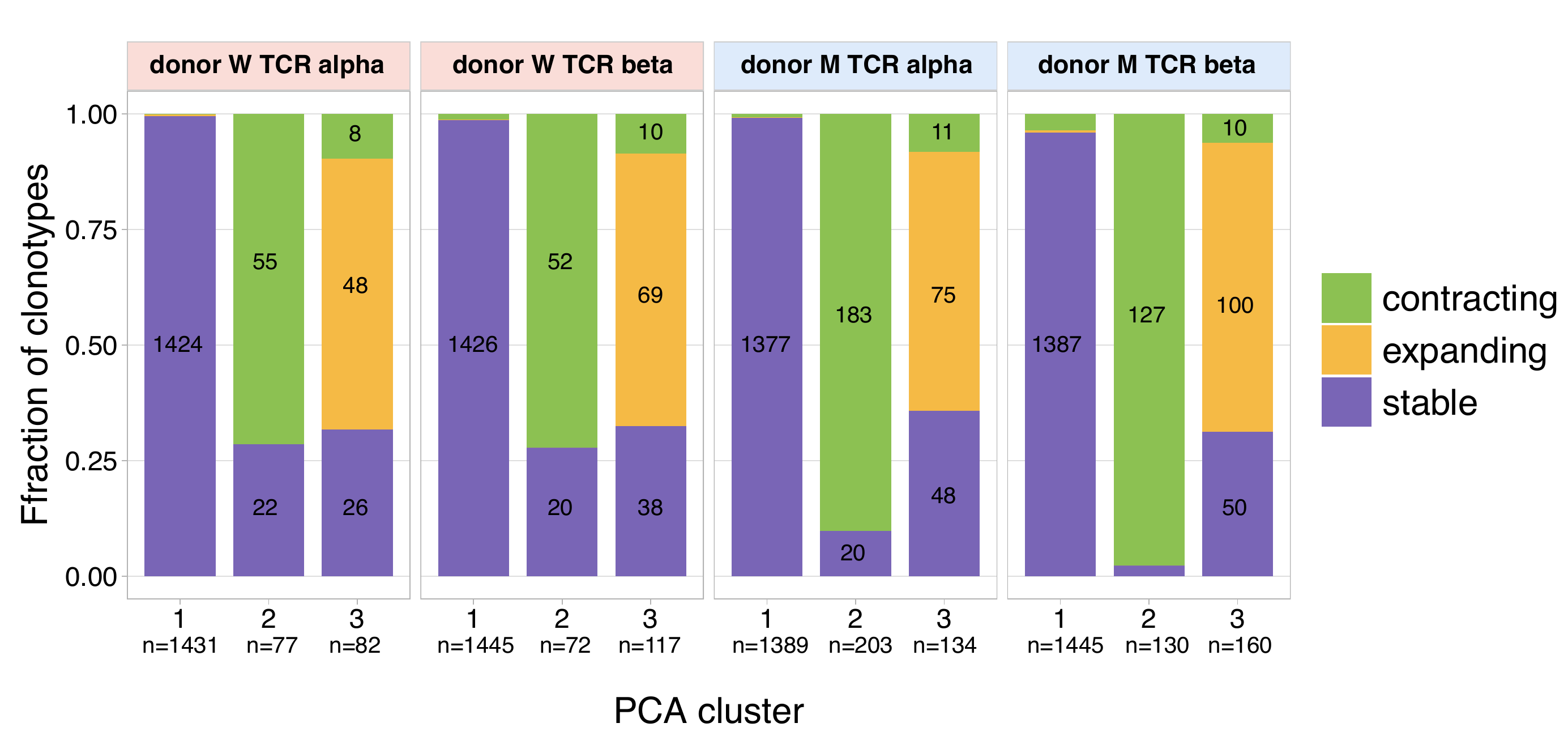}
\caption{{\ch{\bf The overlap between clusters of clonal trajectories identified by PCA and groups of expanding/contracting clones identified with edgeR/NoisET.} For each cluster of clonal trajectories identified on Fig. 1bc. and Fig. S3ab we show overlap with groups of significantly (called by edgeR and NoisET simultaneously) expanding clonotypes from day 15 to day 37 in yellow, significantly contracting clonotypes from day 15 to day 85  in green, other clonotypes are shown in purple ("stable" clonotypes which were not called significant by edgeR and NoisET simultaneously). Bar heights show fraction of abundant clonotypes in PCA cluster overlapping with expanding/contracting/non-significant groups called by edgeR/NoisET, raw number of overlapping clonotypes is shown inside bars. }
}
\end{figure*}

\begin{figure*}[p]
\noindent\includegraphics[width=\linewidth]{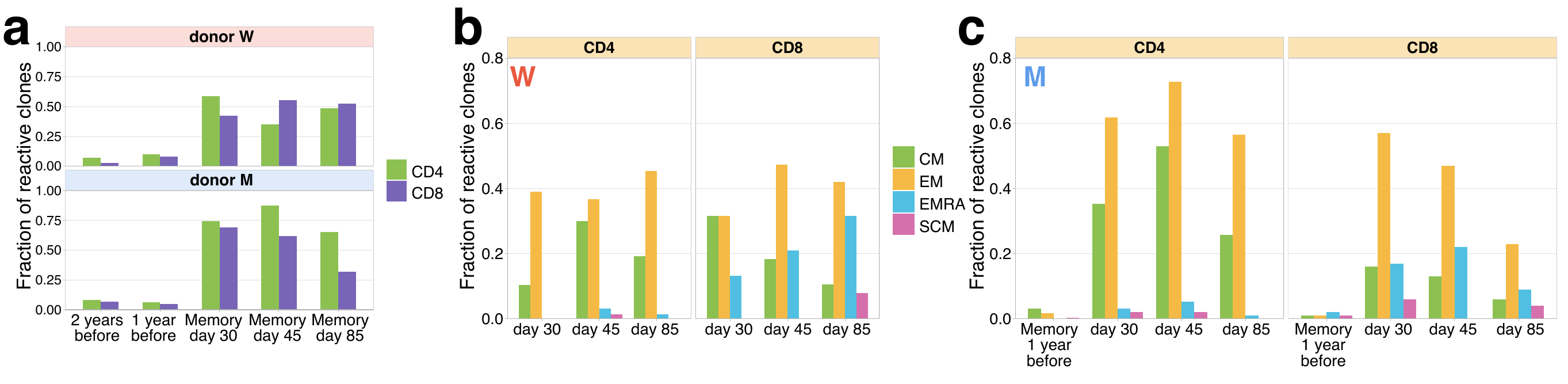}
\caption{{\bf Memory phenotypes of responding clonotypes expanding from day 15 to day 37. a, A fraction of expanding clonotypes is identified in T cell memory subsets after infection.}  Bars show the fraction of expanding CD4+ and CD8+ TCRbeta clonotypes present in 2-year; 1-year pre-infection PBMC; and in at least one of memory subpopulation sampled on day 30 and day 37 post infection. {\bf b, Responding clones are found in different memory subsets.} For both W {\bf(b)} and M {\bf(c)} donors, CD4+ clonotypes were found predominantly in Central Memory (CM) and  Effector Memory (EM), while CD8+ T cells were also present in EMRA. 
}
\end{figure*}

\begin{figure*}[p]
\noindent\includegraphics[width=0.9\linewidth]{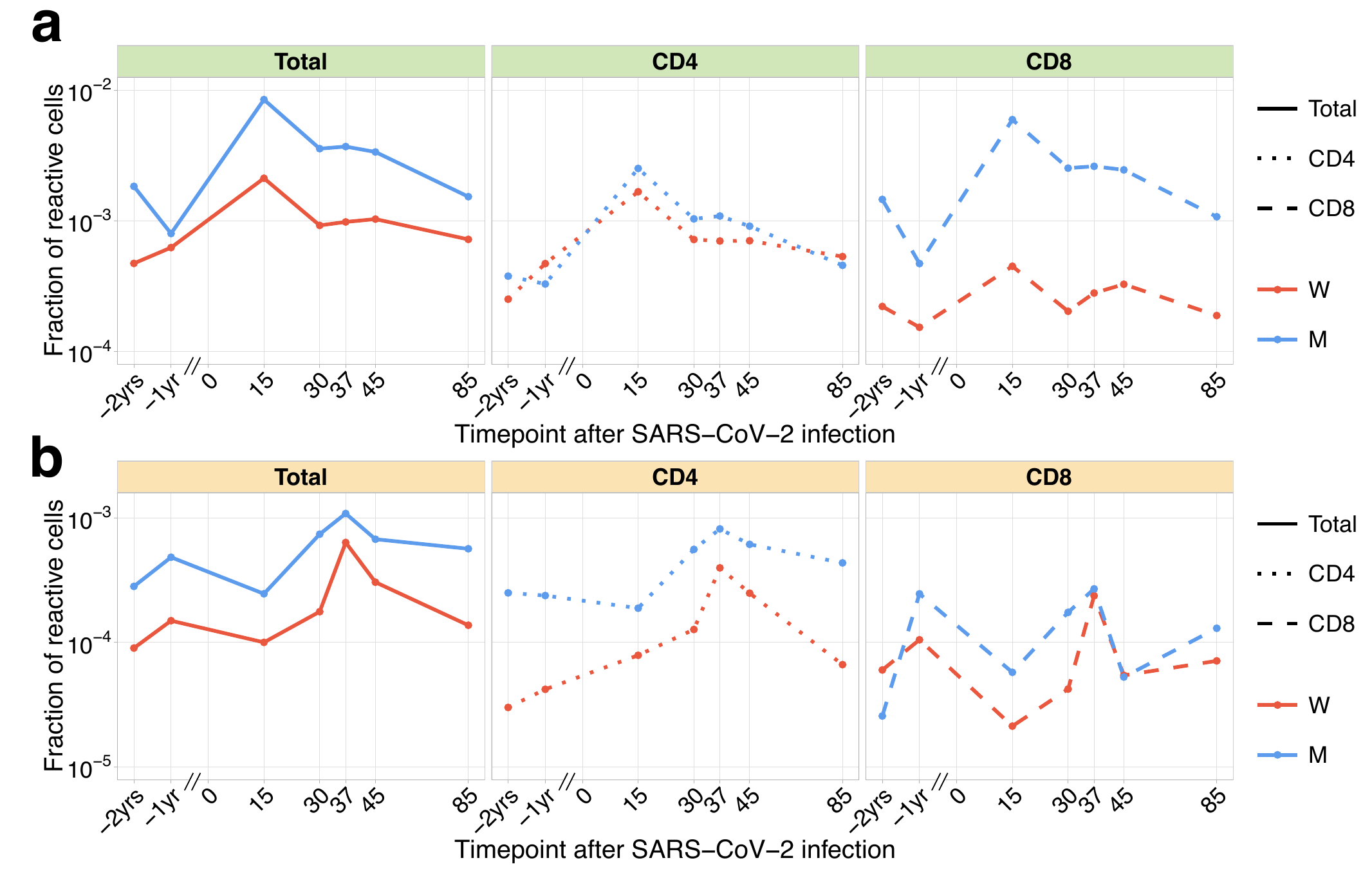}
\caption{{\bf  Dynamics of pre-existing SARS-CoV-2 responding clones.} The fraction of pre-existing (identified in -1 yr and/or -2 yr timepoint pre-infection) contracting {\bf (a)} and expanding {\bf (b)} TCRbeta clonotypes in the total repertoire (corresponding to the fraction of responding cells of all T cells) is plotted in log-scale for all reactive clones (left), reactive clones with the CD4 (middle) and the CD8 phenotype (right). 
}
\end{figure*}

\begin{figure*}[p]
\noindent\includegraphics[width=0.6\linewidth]{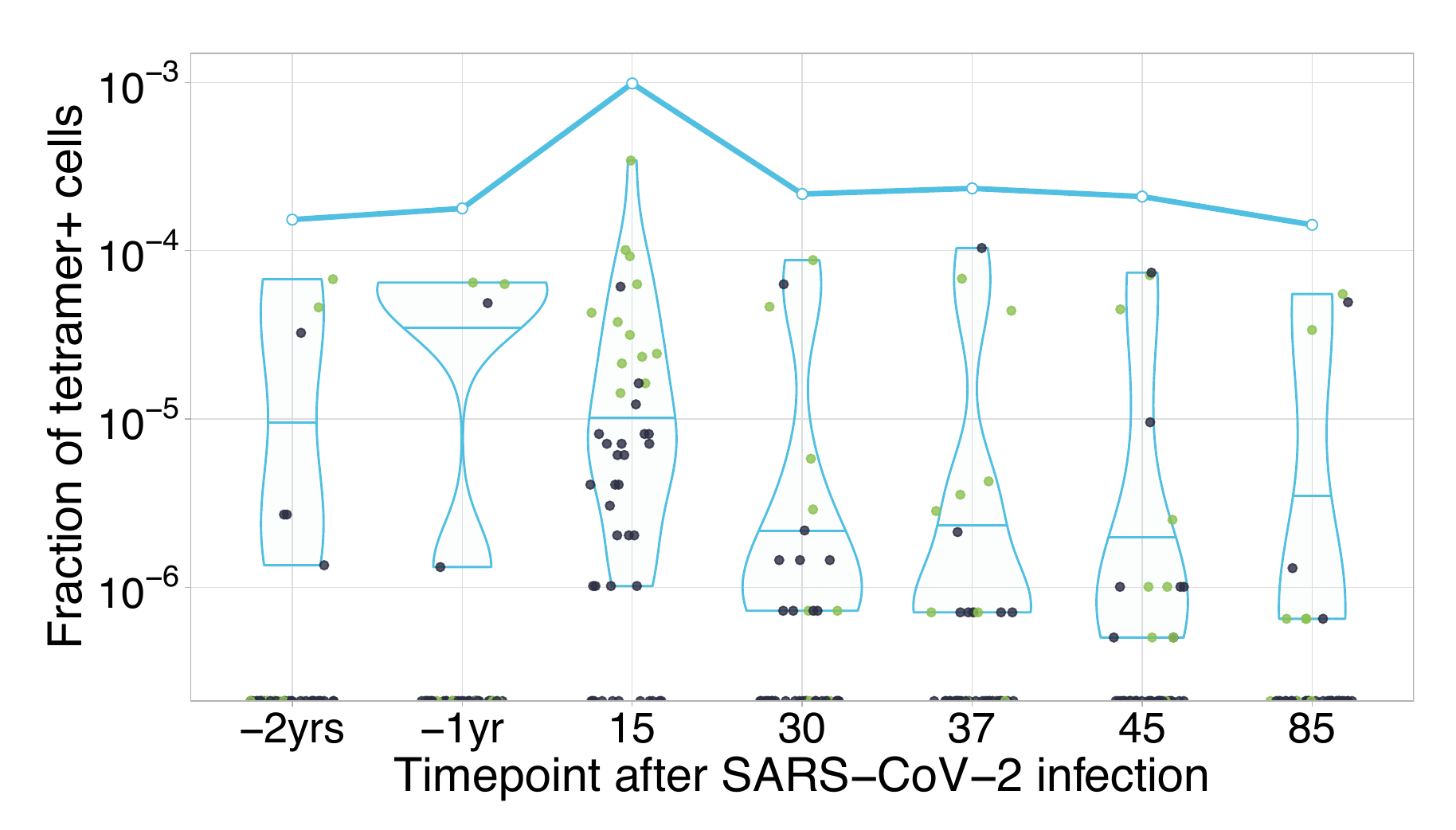}
\caption{{\ch{\bf HLA-A*02:01-YLQPRTFLL-specific TCRs are independently identified by clonal contraction.} Each dot corresponds to the frequency of HLA-A*02:01-YLQPRTFLL-tetramer specific TCRalpha in bulk repertoire from donor M (donor p1434 from \cite{Shomuradova2020}) at given timepoint (an estimate of fraction of tetramer+ cells of all CD3+ cells). Green dots correspond to clonotypes independently identified as contracting in our longitudinal dataset. Blue line shows cumulative frequency of  HLA-A*02:01-YLQPRTFLL-tetramer specific TCRalpha clonotypes.
}}
\end{figure*}

\begin{figure*}[p]
\noindent\includegraphics[width=0.6\linewidth]{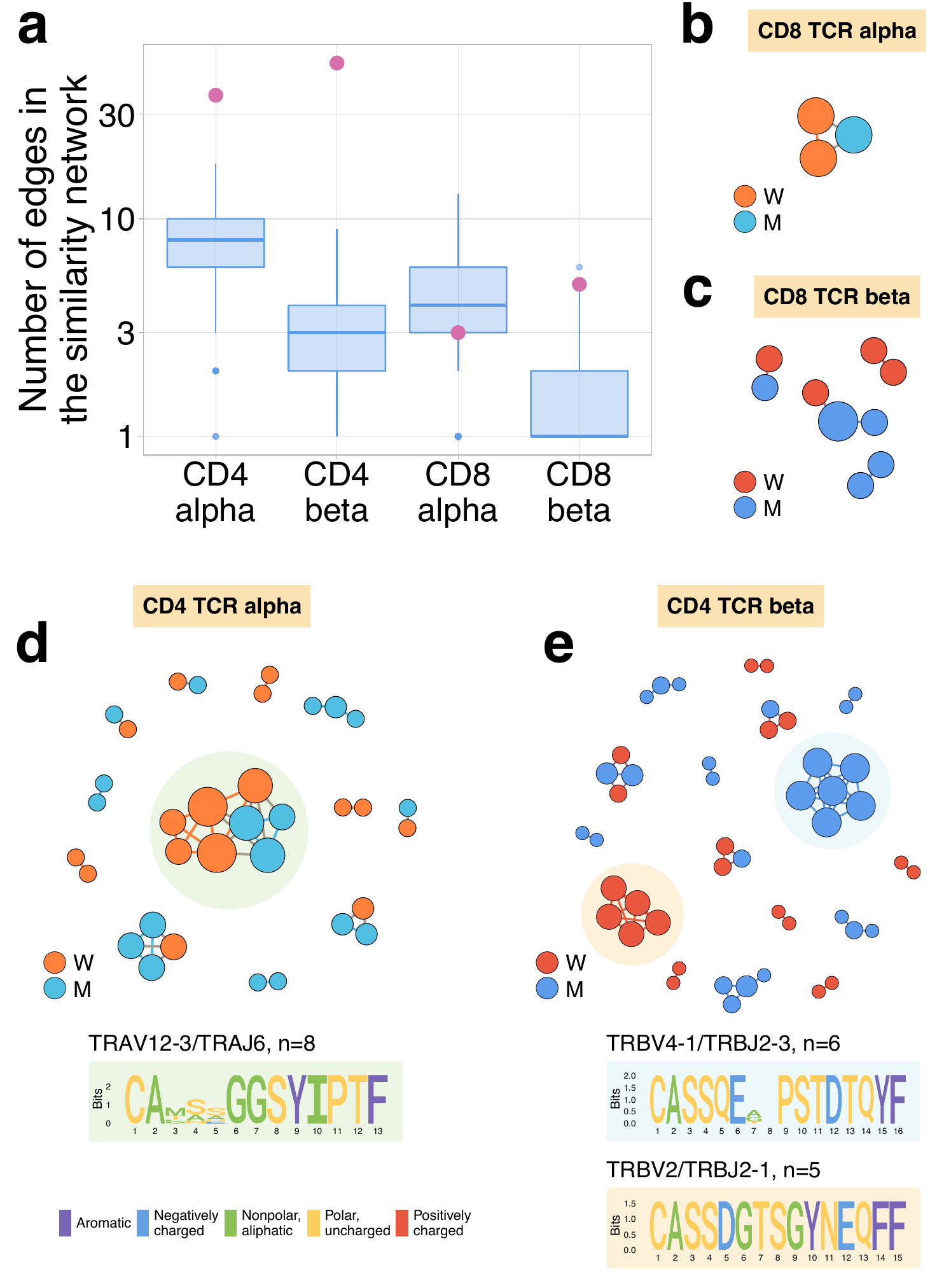}
\caption{ {\ch {\bf a, Expanding CD4+ (but not CD8+) clonotypes show unexpected TCRalpha and TCRbeta sequence convergence.} For each set of CD4alpha, CD4beta, CD8alpha and CD8beta expanded clonotypes, we constructed separate similarity networks. Each vertex in the similarity network corresponds to an expanding clonotype. An edge indicates 2 or less amino acid mismatches in the CDR3 region, and identical V and J segments. The number of edges in each group is shown by pink dots and is compared to the distribution of that number in 1000 random samples of the same size from the relevant repertoires at day 37 (blue boxplots).  {\bf b, c, d, e Analysis of TCR amino acid sequences of expanding clones reveal distinctive motifs.} Networks are plotted separately for CD8alpha {\bf(b)} and CD8beta {\bf(c)} CD4alpha {\bf(d)} and CD4beta {\bf(e)} expanding clonotypes. Clonotypes without neighbours are not shown. Sequence logos corresponding to the largest clusters are shown under the corresponding network plots. }
}
\end{figure*}
\begin{figure*}[p]
\noindent\includegraphics[width=0.65\linewidth]{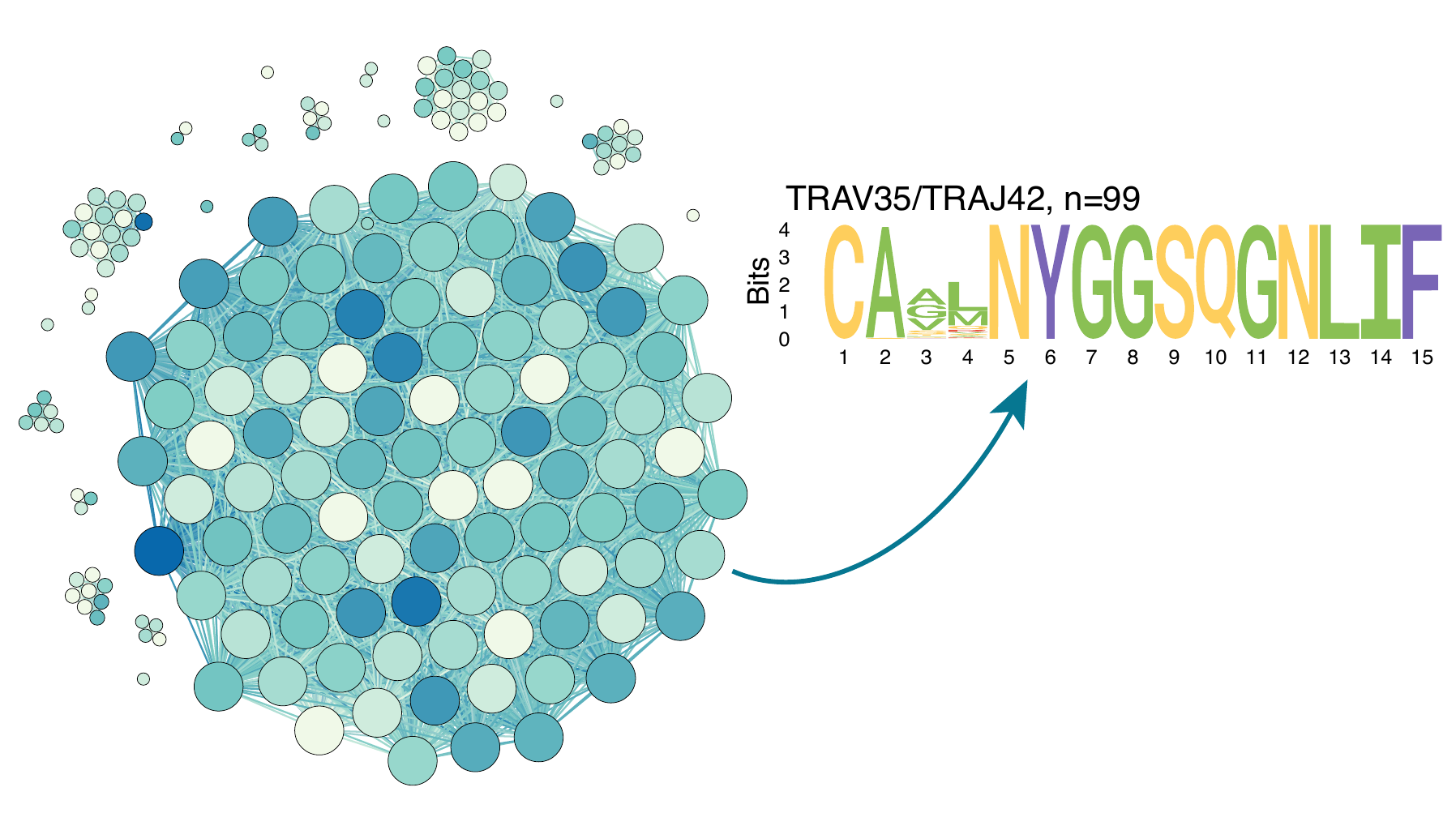}
\caption{{\bf ALICE algorithm output for TCRalpha PBMC repertoire of donor M on day 15.} Similarity network shows ALICE hits (clones in repertoire with more neighbours than expected by chance), which differ by 2 mismatches or less in TCRalpha amino acid sequence. Darker colors indicate larger frequency of clone in the repertoire, vertex size indicates degree. The majority (54\%, 99/183) of hits identified by the algorithm correspond to a single large TRAV35/TRAJ42 cluster of CD4+ contracting clones also seen on Fig. 4a.
}
\end{figure*}

\begin{figure*}[p]
\noindent\includegraphics[width=0.9\linewidth]{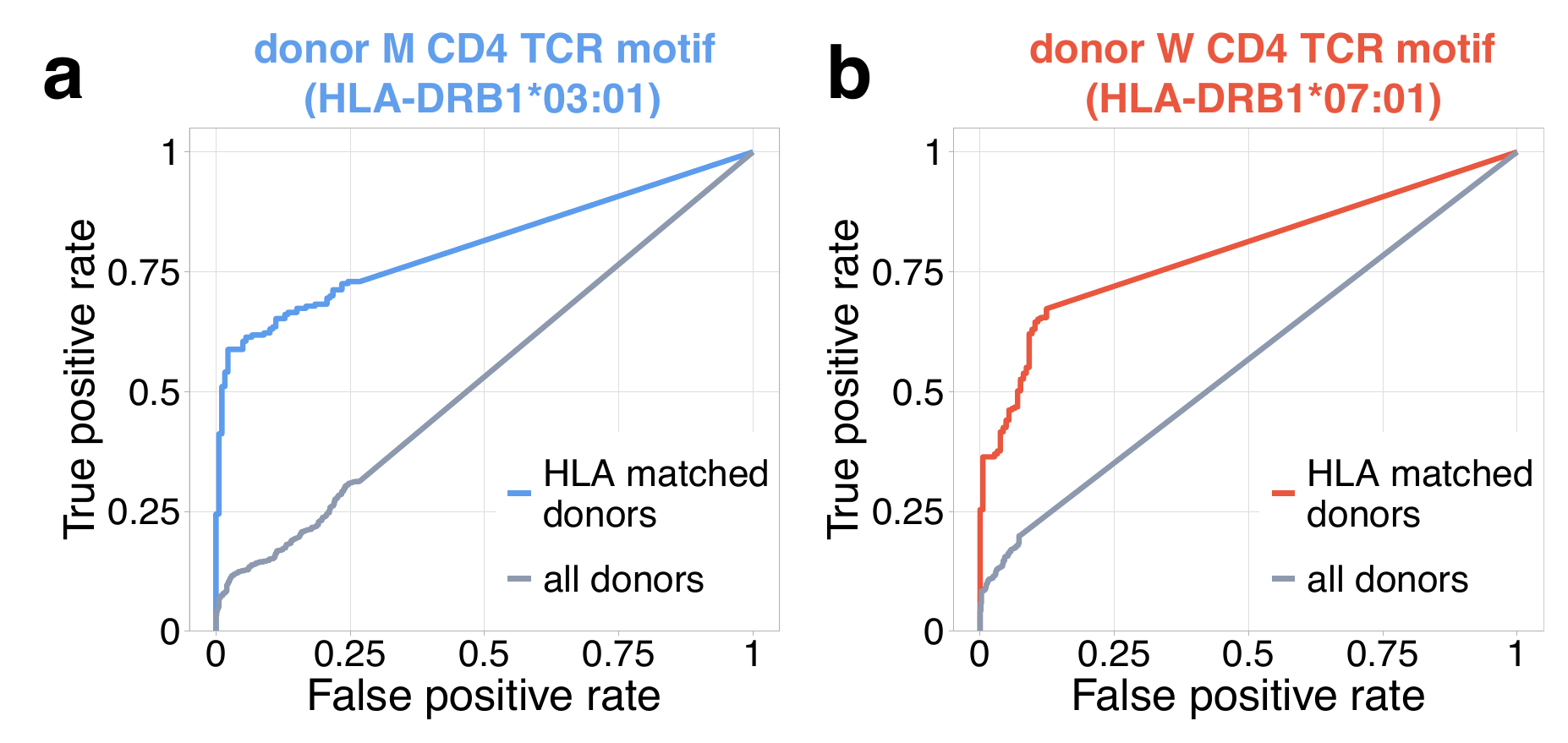}
\caption{{\ch{\bf Identification of COVID-19 patients by frequency of TCR motifs from contracting CD4+ clones from donors M ({\bf a}) and W ({\bf b}).} Receiver Operating Characteristic (ROC) curves for classifying TCRbeta repertoires from COVID cohort vs control by cumulative frequency of clones from CD4beta motifs. Blue curve shows ROC curve (area under the ROC or AUROC=0.8) for the classification of control and COVID donors predicted to be DRB1*03:01-DQB1*02:01 haplotype-positive with motif from donor M. Red curve show ROC curve (AUROC=0.79) for classification of control and COVID donors predicted to be DRB1*07:01-positive using motif from donor W. Grey ROC curves show classifier performance on all donors, irrespective of HLA allele matching (AUROC=0.53 for ({\bf a}), AUROC=0.57 for ({\bf b})). }
}
\end{figure*}

\end{document}